\documentclass[fleqn,usenatbib]{mnras}

\hypersetup{citecolor=blue,urlcolor=magenta}

\usepackage[T1]{fontenc}
\usepackage{booktabs}
\usepackage{graphicx}
\usepackage [caption=false]{subfig}

\DeclareRobustCommand{\VAN}[3]{#2}
\let\VANthebibliography\thebibliography
\def\thebibliography{\DeclareRobustCommand{\VAN}[3]{##3}\VANthebibliography}

\defcitealias{ZZ14}{Z14}

\usepackage{graphicx}	
\usepackage{amsmath}	
\usepackage{amssymb}	

\usepackage{newtxtext,newtxmath}
\usepackage{xcolor}

\title[The relation between AGN outflow and Radio Emission in Radio-Quiet Quasars]{Outflow-related radio emission in radio-quiet quasars}

\author[M. Liao et al.]{
Mai Liao,$^{1,2,3}$\thanks{liaomai@ustc.edu.cn}
Junxian Wang,$^{4,5}$\thanks{jxw@ustc.edu.cn}
Wenke Ren,$^{4,5}$
Minhua Zhou$^{6}$
\\
$^{1}$National Astronomical Observatories, Chinese Academy of Sciences, 20A Datun Road, Chaoyang District, Beijing 100101, China\\
$^{2}$Chinese Academy of Sciences South America Center for Astronomy, National Astronomical Observatories, CAS, Beijing, 100101, China\\
$^{3}$Instituto de Estudios Astrofísicos Facultad de Ingeniería y Ciencias Universidad Diego Portales Av. Ejército 441, Santiago, Chile\\
$^{4}$CAS Key Laboratory for Research in Galaxies and Cosmology, Department of Astronomy, University of Science and Technology of China, Hefei, Anhui 230026, China\\
$^{5}$School of Astronomy and Space Science, University of Science and Technology of China, Hefei 230026, China\\
$^{6}$School of Physical Science and Intelligent Education, Shangrao Normal University, 401 Zhimin Road, Shangrao 334001, China
}

\date{Accepted XXX. Received YYY; in original form ZZZ}

\pubyear{2023}
\begin{document}
\label{firstpage}
\pagerange{\pageref{firstpage}--\pageref{lastpage}}
\maketitle

\begin{abstract}
In this work, we revisit the relationship between [O III] line width $w_{\rm 90}$ (as the indicator of AGN outflow velocity) and the radio emission 
in RQQs by employing a large sample of Type I quasars ($\sim 37,000$) selected from the Sloan Digital Sky Survey (SDSS) Data Release Sixteen. By median stacking the radio images (to include the dominant fraction of individually radio non-detected RQQs) of Karl G. Jansky Very Large Array (VLA) Sky Survey (VLASS) for subsamples of RQQs with different $w_{\rm 90}$, our study demonstrates that, 
the correlation between $w_{\rm 90}$ and radio emission in our SDSS RQQs is significant, and remains solid after controlling the effects of black hole mass, quasar luminosity, Eddington ratio and redshift. This intrinsic link supports that the [O III] outflows in quasars, most likely resulted from wide-angled sub-relativistic quasar winds launched from the accretion disc, could make a dominant contribution to radio emission in the general RQQs. Alternatively,
the correlation may be attributed to low-power jets in RQQs if they are ubiquitous and could efficiently enhance the [O III] width through interacting with the ISM. Meanwhile, the star-formation rates traced by the flux ratio of [Ne V]/[O II] emission lines 
display no dependence on $w_{\rm 90}$ after controlling the effects of black hole mass, quasar luminosity, Eddington ratio and redshift. This suggests that the stronger radio emission in RQQs with larger $w_{\rm 90}$ could not be attributed to outflow enhanced (positive feedback) star formation in the hosts. However, this also indicates the outflows, though exhibiting robust correlation with radio power, produce neither positive nor negative feedback to the star formation in their hosts.

\end{abstract}




\section{Introduction}
Radio-quiet quasars (RQQs) make up the majority of active galactic nuclei (AGN) population with a fraction of $\sim 90\%$, whereas only $\sim 10\%$ quasars are radio-loud (RLQs). The taxonomy of these two populations, traditionally, is based on the radio loudness which is defined as the ratio between radio and optical flux (e.g., $R = f_{\rm 5GHz}/f_{\rm 4400\AA}$, \citealt{1989AJ.....98.1195K}). But see \cite{Padovani2017NA} for an alternative naming of RL/RQ as jetted and non-jetted.
Compared with the RLQs (e.g., M87 and Cygnus A) where the relativistic and powerful jets dominate the radio emission \citep[e.g.,][]{1995PASP..107..803U,2019ARA&A..57..467B,2020NewAR..8801539H}, however, the physical mechanism responsible for the radio emission in radio-quiet ones (not radio-silent) is still not fully understood, as the emission is on average 1000 times fainter than RLQs and also compact which is harder to spatially resolve \citep[e.g.,][]{1993MNRAS.263..425M,2019NatAs...3..387P}. There are several proposed mechanisms which include low-power compact jets, magnetically heated accretion-disc corona, outflows from disc winds, and star formation of the host galaxies \citep[e.g.,][]{2019NatAs...3..387P,2023Galax..11...27K,2023Galax..11...85S,2023ApJS..269...24K}. Understanding the physical origin of the radio emission in RQQs, is crucial to our understanding of the complete picture of the AGN phenomenon, including AGN accretion and feedback (radiative mode/quasar wind versus mechanical mode of the radio jet).

Observations on morphology are the most direct way to identify the origin of radio emission of RQQs. For example, host-like, diffuse and/or clumpy structures likely indicate star formation process; the biconical emission perhaps originates from outflow; and the clear detection of core-jet like structures would be a strong evidence of the presence of relativistic jet \citep{2019NatAs...3..387P}. However, the radio emission in RQQs and RQ AGN is predominantly compact/unresolved at (sub-)arcsec scales \citep[e.g.,][]{2010AJ....139.1089L,2021AJ....162..270R,2020MNRAS.499.5826S,2021MNRAS.503.1780J,2022MNRAS.515..473P,2022AJ....164..122M}, and even at milliarcsecond scales \citep[e.g.,][]{1998MNRAS.299..165B,2005ApJ...621..123U,2016A&A...589L...2H,2022ApJ...936...73A,2023MNRAS.518...39W}, keeping the emission mechanism elusive, even though structures of double/triple/jet-like/
diffuse features have been resolved in a number of objects with the currently known but limited number of Very Large Array (VLA) and Very Long Baseline Interferometry (VLBI) observations.

Quasar outflows are prevalent and widespread throughout AGN \citep[e.g.,][]{2012ApJ...745L..34Z,2014MNRAS.441.3306H}. They are routinely invoked as energetic feedback from AGN to their host galaxies to explain the scaling relations between the mass of the central black holes and their host galaxy properties \citep[e.g.,][]{2008ApJS..175..356H,2013ARA&A..51..511K}.
The subrelativistic outflows ($\geq$ 0.1 c) around black hole at sub-pc scale have been revealed by the observations of X-ray and ultraviolet spectroscopy \citep[e.g.,][]{2003ARA&A..41..117C,2003ApJ...593L..65R,2010A&A...521A..57T,2011ApJ...742...44T,2013MNRAS.430.1102T}. At galaxy scales of (sub-) kpc, the ionized gas outflows with moderate velocity, traced by the broad and asymmetric [O III] $\lambda 5007$ (shortly [O III] hereafter) emission line profile \citep{2006agna.book.....O}, have been commonly found 
\citep[e.g][]{2013MNRAS.433..622M,ZZ14,2018ApJ...865....5R}. The quasar-driven disc winds \citep{2007ASPC..373..267P} are usually proposed to be the driving mechanism for these outflows in RQQs via interacting with the surrounding gas at various spatial scales. Meanwhile, the generated shock in such interaction, could accelerate the electrons and produce synchrotron emission \citep[e.g.,][]{2010ApJ...717..708C,2010ApJ...711..125J,2015MNRAS.447.3612N}.
As manifested from the modelling for such radio emission of RQQs in \cite{2015MNRAS.447.3612N}, the proportion of the synchrotron emission to bolometric luminosity is about $10^{-6} -$ $10^{-5}$ if 0.5-5\% of the bolometric luminosity is transferred to the kinetic energy of the outflow, and then $\sim 0.01\%$ outflow energy is used to accelerate electrons to produce the radio emission. 

It would imply that the radio emission would correlate with the outflow properties if both are driven by the nuclear quasar powered winds. Historical researches suggested that [O III]-traced outflows and the extended radio emission are co-spatial, implying a shock-related origin of the radio radiation \citep[e.g.,][]{1997MNRAS.284..385C}.
Furthermore, 
the correlation between the width of [O III] and radio luminosity has been found in literatures for radio-quiet AGN and RQQs with the kinematics of [O III] emission line dominated by outflows \citep[e.g.,][]{1991ApJS...75..383V, 2013MNRAS.433..622M,ZZ14,2018MNRAS.477..830H}. 
In particular, \cite{ZZ14} (hereafter \citetalias{ZZ14})
found that the [O III] line width, i.e., $w_{\rm 90}$ which serves as a proxy of the outflow, strongly correlates with the radio luminosity among 568 luminous obscured Type II low-redshift ($z$ < 0.8) quasars, indicating that the radio emission in RQQs could be due to relativistic electrons accelerated in the shocks within the quasar-driven outflows. Moreover, the extremely red radio-quiet quasars of \cite{2018MNRAS.477..830H} at high-redshift lie on the extension of the relation in \citetalias{ZZ14}. 

Note the advantage of using the width [O III] line of $w_{\rm 90}$ to trace the outflow strength, is that it's sensitive to the presence of a broad wing component, i.e., the Doppler broadening from the motion of the outflow gas, but not affected by the outflow geometry. Although the blueshifted broad wing component, which is deemed as a 'smoking gun' of outflow signature, if under the case of spherically symmetric or biconical outflow, the velocity shift relative the systematic velocity around zero and would  not display any blueshift on the [O III] line profile \citep[e.g.,][]{2013MNRAS.436.2576L,ZZ14,2016ApJ...817..108W,2018ApJ...865....5R}.
The stronger outflows would cause more intense turbulence on outflows gas and hence increase the width of the [O III] emission line.
Therefore, $w_{\rm 90}$ can be used as the indicator of outflow strength regardless of the profile of [O III] and utilizing $w_{\rm 90}$ allows us to construct of much larger optical sample to explore the shocked radio emission caused by NLR outflows as opposed to [O III] blueshift.

In this study, we revisit the 
 correlation between [O III] $w_{\rm 90}$ and radio emission,
using a type 1 RQQ sample selected from SDSS DR16, specifically through controlling the effects of other parameters (i.e., redshift $z$, black hole mass $M_{\rm BH}$, quasar luminosity $L_{\rm bol}$, Eddington ratio $R_{\rm Edd}$) which may lie behind such correlation. We take radio non-detection sources into consideration and derive their average radio emission by stacking the images of Karl G. Jansky Very Large Array (VLA) Sky Survey \citep[VLASS,][]{2020PASP..132c5001L} which completely cover the SDSS survey area. Our work allows us to explore the intrinsic relation between $w_{\rm 90}$ and radio emission in radio non-detection regime which is not exploited yet before,
thus could derive more general pattern since radio non-detection ones represent the main population of RQQs, and help to better understand the shocked-related radio emission. Our paper is structured as follows: \S2 presents our optical sample selection and spectroscopic analysis; \S3 shows the stacking analyses of the radio images for the optical sample, and the derived results; 
\S4 provides a discussion. Throughout the paper, the cosmological parameters $H_0 = 70\, \mathrm{km\,s^{-1}\, Mpc^{-1}}$, $\Omega_\mathrm{m} = 0.3$, and $\Omega_{\lambda} = 0.7$ are adopted. 

\section{The quasar sample}
The quasar sample presented in this work is constructed from the Sloan Digital Sky Survey (SDSS) Data Release sixteenth Quasar Catalog \citep[DR16Q,][]{2020ApJS..250....8L}. It is selected based on the quasars' optical spectroscopic properties, namely, with significant [O III] emission line, and the measurements of $M_{\rm BH}$, $L_{\rm bol}$, $R_{\rm Edd}$. In the following, we describe the detailed process of sample construction.



SDSS DR16Q consists of 750,141 quasars accumulated from SDSS-I to SDSS-IV. The spectra from SDSS-I/II have a wavelength coverage from 3800 to 9200 \AA~with a spectral resolution R of 2000, while 3600 -- 10400 \AA~at the resolution of 1300-2500 for SDSS-III/IV.
We focus on quasars with redshift $z$ < 1.0 for SDSS-III/IV, and $<$ 0.77 for SDSS-I/II, to guarantee the [O III] line coverage of the spectra.
Simultaneously a median signal-to-noise (S/N) ratio $>5$ $\rm pixel^{-1}$ over the whole spectrum is applied. This threshold is lower than those adopted in some previous works \citep[e.g., S/N > 10 $\rm pixel^{-1}$ in ][]{2018ApJ...865....5R},  
to allow retention of quasars with strong [O III] emission line but low continuum S/N. 
The local median S/N $\geq 5$ $\rm pixel^{-1}$ in the line-fitting region of $\rm H\beta$-[O III] (see Section 2.1 below) are further imposed, to ensure reliable measurements \citep{2011ApJS..194...45S} for [O III] and broad $\rm H\beta$ line, the latter of which will be used to derive $M_{\rm BH}$. 
These steps result in 51891 targets. 

\subsection{Optical spectroscopic analysis}
We analyze SDSS spectra adopting the publicly available Python code PyQSOFit\footnote{https://github.com/legolason/PyQSOFit}
to fit the continuum and emission lines \citep{2018ascl.soft09008G}. 
The SDSS spectra are first corrected for Galactic extinction \citep{1998ApJ...500..525S}, and shifted to the rest-frame wavelength based on the redshifts from DR16Q. For each spectrum, we perform continuum fitting over two spectral ranges separately (local fitting, as discussed in \citealt{2011ApJS..194...45S}), i.e., 3540-6900 \AA~around H$\beta$ and H$\alpha$, and 2200-3100 \AA~around Mg II.  
For each spectral range, the continuum  in line-free regions is modelled with PyQSOFit as the linear combination of a powerlaw (AGN continuum) and Fe II emission.
For the spectral range of 3540-6900 \AA, a host galaxy component (but only for sources with clear absorption lines of Ca H $\&$ K, G band, Mg b and NaD  detected, see \citealt{2020MNRAS.491...92L}) is also included.
The continuum luminosity at 3000 and 5100~\AA~are then derived using the power-law model. 

We fitted the emission lines using Gaussian profiles on the
continuum-subtracted spectra utilizing line-fitting windows the same as those adopted in \cite{2011ApJS..194...45S}. In order to select the quasars with significant [O III] emission line, each of the [O III] $\lambda\lambda4959,5007$ doublets are firstly fitted using a single Gaussian with an upper limit of 1200 $\rm km~ s^{-1}$ for line FWHM, where the line width and velocity offset are linked for the doublet and the flux ratio fixed to the theoretical value of 3. The ones with (S/N)$_{\rm line}$ of [O III] > 3 are chosen to further analyses, where (S/N)$_{\rm line}$ is defined as the ratio of the peak amplitude of [O III] 5007 to the noise that is the standard deviation calculated over the spectral window of 5100$-$5200 \AA\ after subtracting the fitting models of continuum and emission line. After this criterion, we obtain 39777 quasars. For these objects, their [O III] $\lambda\lambda4959,5007$ emission lines are further fitted by adding a second broad Gaussian component to represent potential asymmetric blue or red wings. For the final [O III] profile, only if the second Gaussian component had (S/N)$_{\rm line}$ > 3 the double Gaussian profile is adopted, otherwise the original single Gaussian profile is utilized. 

The broad components of H$\beta$, H$\alpha$ (when available), and Mg II are fitted with one or multiple Gaussian profiles (up to three) with FWHM > 1200 $\rm km~ s^{-1}$ \citep{2011ApJS..194...45S}. 
Additionally, all the narrow line components of H$ \beta$, H$\alpha $, [N II] $\lambda\lambda6548,6584$, [S II] $\lambda\lambda6716,6731$ and Mg II are modelled with a single Gaussian with FWHM < 1200 $\rm km~ s^{-1}$. The line width and velocity offset for narrow H$\alpha $, [N II] $\lambda\lambda6548,6584$, [S II] $\lambda\lambda6716,6731$ are linked, and the flux ratio of [N II] doublet fixed at 2.96.

We further limit our analyses to 37,218 quasars with the detection of at least one broad line of H$\beta$, Mg II, or H$\alpha$ (S/N$_{\rm line}$ >3 as did for [O III]), which will be adopted for the BH mass and Eddington ratio estimation. Similarly, the median S/N in the line fitting region of Mg II or H$\alpha$ is required to $\geq 5$ $\rm pixel^{-1}$, as the precondition of corresponding line fitting. 


To assess the uncertainties in the measured quantities in our fitting, we generate 50 mock spectra by adding random Gaussian noise (based on the observed flux density errors) to each original spectrum, fit those mock spectra following the same routines, and calculate the standard deviation of each measured quantity from the mock spectra. 

\subsection{Black hole mass, bolometric luminosity and Eddington ratio estimation}
We derive the $M_{\rm BH}$ of our quasars from the
single-epoch SDSS optical spectra assuming the broad line region (BLR) is virialized:
\begin{equation}
\log \left(\frac{M_{\rm BH}}{\rm M_{\rm \sun}} \right)=a+b\log\left(\frac{\lambda L_{\rm \lambda}}{10^{44}\
\rm ergs\ s^{-1}} \right) +2\log\left(\frac{\rm FWHM}{ \rm km\ s^{-1}} \right)
\end{equation}
where $\lambda L_{\rm \lambda}$ is the continuum luminosity at 5100 \AA~for H$ \beta$ line, and 3000 \AA~for Mg II, respectively. The coefficients a and b are the calibration parameters, taken from \cite{2006ApJ...641..689V} for H$ \beta$ of (a, b) as (0.91, 0.5) and \cite{2009ApJ...699..800V} for Mg II as (0.86, 0.5). We prefer H$\beta$ based $M_{\rm BH}$ over Mg II if both are available. In the cases of both broad H$\beta$ and Mg II being unavailable, either statistically non-detected or the median S/N $\rm pixel^{-1}$ around line fitting region smaller than 5, we use the alternative empirical recipe of equation (8) in \cite{2011ApJS..194...45S} which employs the luminosity and FWHM of $\rm H\alpha$ to estimate the $M_{\rm BH}$ if $\rm H\alpha$ available.
We uniformly calculate the $L_{\rm bol}$ from $L_{\rm 5100}$ for all spectra using the corrections $\rm BC_{\rm 5100}$ = 9.26 \citep{2011ApJS..194...45S}. 
The Eddington ratio $R_{\rm Edd}= L_{\rm bol}/ L_{\rm edd}$ can then be calculated where $L_{\rm edd} = 1.38 \times 10^{38} M_{\rm BH}/\rm M_{\sun}\ erg\ s^{-1}$. 
The measured log $M_{\rm BH}$ ($\rm M_{\rm \odot}$), log $L_{\rm bol}$ ($\rm erg~s^{-1}$) and log $R_{\rm Edd}$ range from 6.6 -- 10.4 with a median of 8.4, 41.70 -- 47.28 with a median of 45.40, and -4.15 -- 0.14 with a median of -1.14, respectively. We note 82\% of our RQQs have log $L_{\rm bol}$ $\geq$ 45, and 98\% of RQQs have log $R_{\rm Edd}$ $>$ -2, i.e.,  usually supposed to be powered by  optically thick standard accretion disc (the high-accretion mode), while only a minor fraction of them could be driven by optically thin advection dominated accretion flow (i.e., low-accretion mode, \citealt{1995ApJ...452..710N}).

In summary, our final quasar sample is formed by 37218 quasars at $z$ < 1, with [O III] $\lambda5007$ detection as well as the measurements of SMBH mass, bolometric luminosity and Eddington ratio. 

\subsection{$w_{\rm 90}$ estimation and subsamples with various $w_{\rm 90}$}
The $w_{\rm 90}$, a non-parametric characterization for the [O III] width and an indication of outflow kinematics, is determined with the best-fit model of the emission-line profile (Section 2.1). Following \citetalias{ZZ14}, $w_{\rm 90}$ 
is defined as $w_{\rm 90}$ = $\upsilon_{95} - \upsilon_{05}$, where $\upsilon_{05}$ and $\upsilon_{95}$ are the
velocities at which 5\% and
95\% of the line flux accumulates, respectively. For a single Gaussian profile, the value of w90 equals to 1.4$\times$FWHM or 3.29$\sigma$. We find the measured $w_{\rm 90}$ in our sample ranging from 250 to 4200 $\rm km~\rm s^{-1}$,  with a median value of 820 $\rm km~\rm s^{-1}$ which is slightly smaller than that of \citetalias{ZZ14} (1060 $\rm km~\rm s^{-1}$). The larger $w_{\rm 90}$ of \citetalias{ZZ14} can be understood that their sources are on average more luminous (with an average log $L_{\rm [O III]}$ of 42.48, compared with 42.10 of our sample), thus stronger winds/radiative-force are expected.

In order to study the relation between radio emission and $w_{\rm 90}$ in RQQs, we divide our quasar sample equally into seven subsamples (referred to as $W_{1}$ to $W_7$ hereafter) according to the ranked $w_{\rm 90}$.
However, both radio emission and $w_{\rm 90}$ could depend on the fundamental AGN parameters including $z$, $M_{\rm BH}$, $L_{\rm bol}$, and $R_{\rm edd}$, which may lie behind the observed correlation between radio emission and $w_{\rm 90}$ and yield biased link between them. 
For instance, with increasing $w_{\rm 90}$, the average of $z$, $M_{\rm BH}$, $L_{\rm bol}$ and $R_{\rm edd}$ of each subsamples do increase.   
Thus, on the grounds of this, these fundamental parameters should be controlled to reveal the intrinsic connection between the shock contributed radio emission and $w_{\rm 90}$. For each subsample we build a control sample with identical size 
and the fundamental parameters aforementioned matched between them. The control sample is derived from the parent sample but excluding the corresponding subsample. 
For instance, the control sample for $W_{\rm 7}$ (which has the largest $w_{\rm 90}$) is selected from $W_{\rm 1}$ to $W_{\rm 6}$.
We adopt one-to-one match by finding for each quasar a control quasar with the minimum distance in the space of $z$, $M_{\rm BH}$, $L_{\rm bol}$, and $R_{\rm edd}$.
As expected, each subsample and its control sample show statistically consistent distributions in these physical parameters (confirmed with the Kolmogorov$-$Smirnov test) . We name the control samples as $MW_{1}$ to $MW_{7}$, respectively.

\section{Radio images and stacking}

VLASS is conducted with NRAO Very Large Array (VLA) in its B-configuration centered at 3 GHz (2 $-$ 4 GHz) and covers the whole sky visible to the VLA (Declination > - 40 deg) with a total sky coverage of about 33,885 square degrees. 
The VLASS images have the angular resolution of 2.5\arcsec, and a typical rms sensitivity of $\sim$ 120 uJy\footnote{1$\sigma$ sensitivity of each VLASS epoch images. The coadded of images taken from three-epoch VLASS observations will achieve an 1$\sigma$ sensitivity of 70 uJy.}. Compared with the VLA Faint Images of the Radio Sky at Twenty-Centimeters (FIRST) survey \citep{1995ApJ...450..559B,2015ApJ...801...26H}, which
works at 1.4 GHz with B-configuration and covers $\sim$ 10575 deg$^{2}$ with a similar typical sensitivity of 150 uJy,
VLASS has the advantage of larger survey area which overlays the entire sky coverage of SDSS, thus allows us to build a larger RQQs sample.

Of our [O III] quasar sample where all the objects are located within the VLASS survey footprint, we find VLASS counterparts for 6.6\% of them through matching with the VLASS catalogue \citep{2020RNAAS...4..175G,2021ApJS..255...30G} with a matching radius of 2.5$''$. In the VLASS detected sources, 72$\%$ can be classified as radio-loud based on their radio-loudness. It indicates that while the dominant population of our sample is radio-quiet, some (5$\%$, assuming 10\% of quasars are radio-loud) radio non-detected sources could still be radio-loud \citep{2022MNRAS.512..296L}.

The stacking technique has the power to measure the average flux of quasars that are individually below the nominal detection threshold (i.e., undetected) in a particular survey. Radio-stacking has been widely employed in a number of studies for large sample of RQQs \citep[e.g.,][]{2005MNRAS.360..453W,2007ApJ...654...99W,2019MNRAS.490.2542P,2022MNRAS.512..296L}.
Below we briefly explain our procedures to derive the VLASS stacked images of our subsamples of RQQs, following the analysis of \cite{2022MNRAS.512..296L} which stacked the FIRST images of RQQs.
For a more detailed explanation of the procedures, we refer the reader to the Section 2.3 of \cite{2022MNRAS.512..296L}.
We employ the median stacking approach which is insensitive to outliers (i.e., the tails of the underlying flux distribution, e.g., less sensitive to the small population of radio-loud sources) and more robust for non-Gaussian distributions when compared with mean stacking \citep{2007ApJ...654...99W}. 

The radio cutout image of VLASS Quick Look (QL) for each quasar, centred on the SDSS quasar position with size of $11'' \times 11''$ and pixel size of $1''$, is extracted from the VLASS survey. VLASS QL images are now available from the VLASS archive image cache for the whole of Epochs 1 and 2 (divided into the VLASS1.1, VLASS1.2, VLASS2.1 and VLASS2.2 campaigns). In this work, we use the cutout images of Epoch 2. We firstly align and median stack the VLASS QL images on a pixel-by-pixel basis, and the average radio flux density of a sample is derived from the stacked image. As a point-like source is centred at the position of
the nominal quasar(s) in each co-added image (see Section 4), and the changing PSFs from various individual VLASS QL images without source detections are not CLEANed, the peak flux density only from the central pixel in each co-added image is adopted as the average radio flux density for later statistic analysis. Since the VLASS survey is not sufficiently deep to detect or exclude all radio-loud quasars from our samples, when performing median-stacking we derive the 45th percentile for all quasars in each subsample (to minimize the effect of $\sim$ 10\% radio loud ones, \citealt{2022MNRAS.512..296L}), instead of the simple median (50th percentile) for all radio non-detected quasars.
Hereafter, if not otherwise specified, we refer to the 45th percentile as the median for RQQs in each $w_{\rm 90}$ subsample.

\begin{figure}
 \centering
	\includegraphics[width=2.9in]{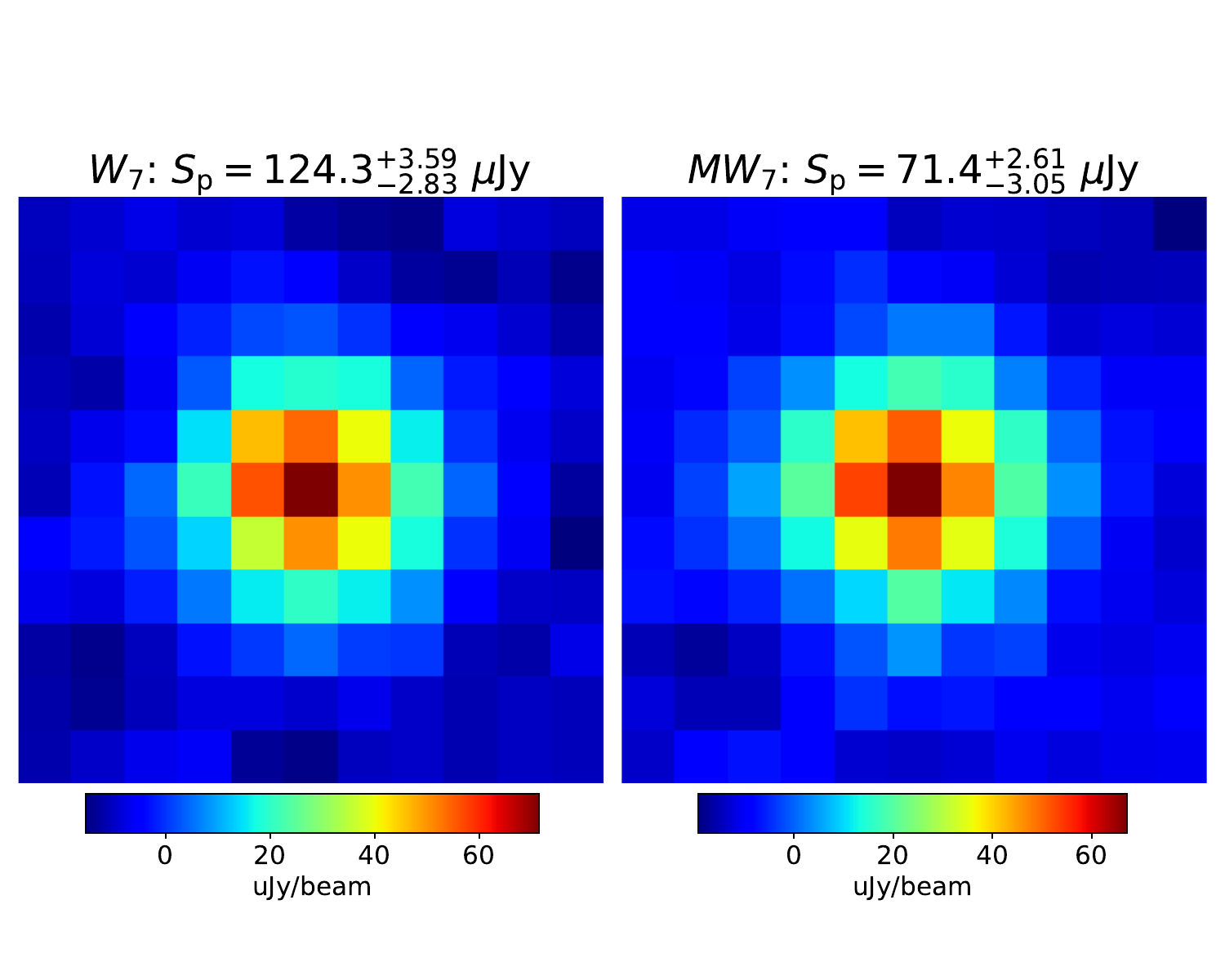}
	\caption{The example (subsample $W_{\rm 7}$) of coadded 3 GHz VLASS images of RQQs (left) and its corresponding control sample (right). The peak flux densities
indicated by the central pixel in each stacked image are shown at the top of
each image. All the coadded images have a scale of 11 $\times$ 11 arcsec$^{2}$ with
pixel size 1 arcsec.}
\label{fig:coaddimage}
\end{figure} 



We show the example of our coadded VLASS 3 GHz images of subsample $W_{\rm 7}$ and its control sample $MW_{\rm 7}$ in Fig. \ref{fig:coaddimage}. Similarly, significant and compact radio signals are detected in the median-stacked images of all other subsamples and corresponding control samples,
certifying that the stacking approach is capable of detecting the average radio emission from our large sample of RQQs. The asymmetric error bars of each peak flux density are obtained by following the method of bootstrapping  described by \citep{2011ApJ...730...61K}.

\begin{figure*}
	\includegraphics[width=3.4in]{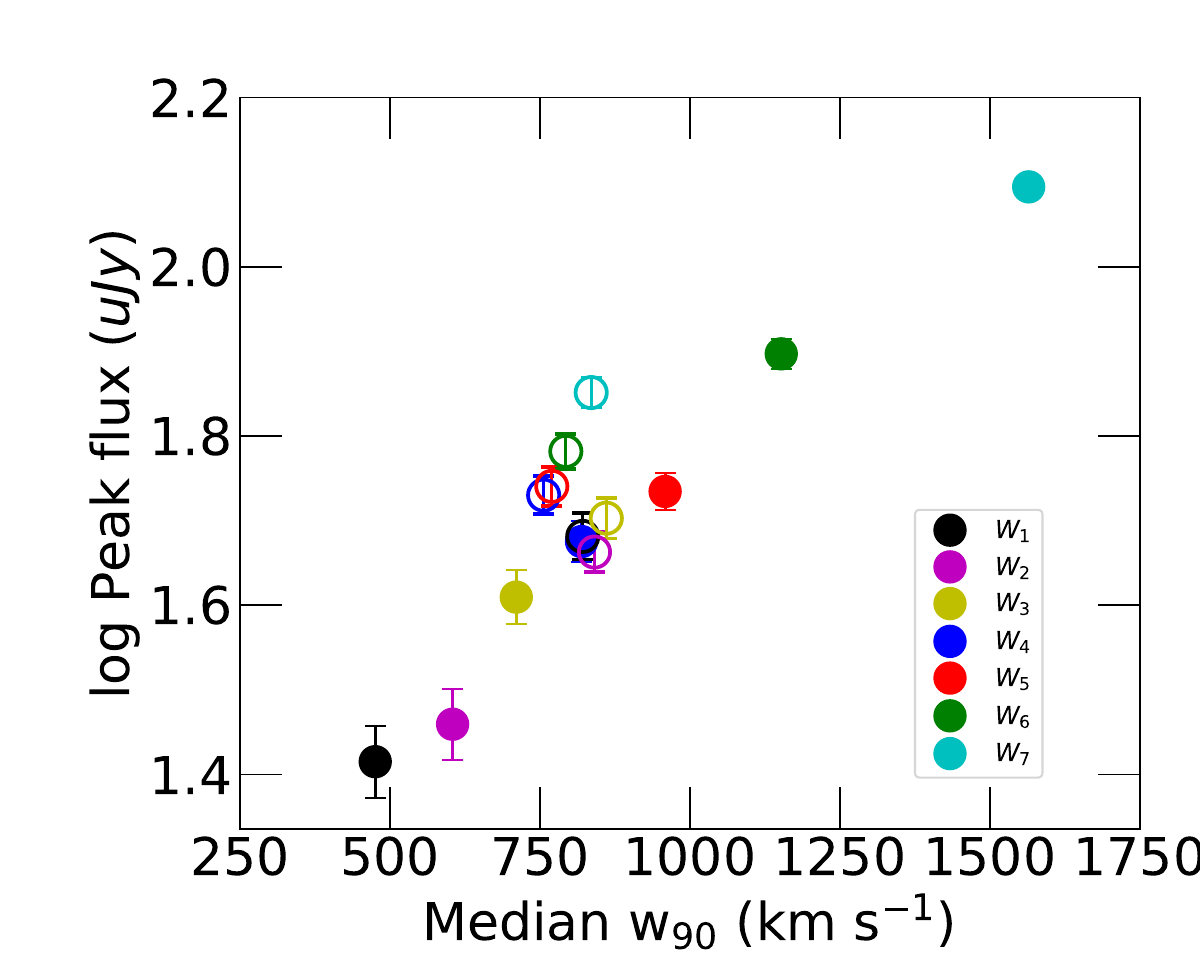}
	\includegraphics[width=3.4in]{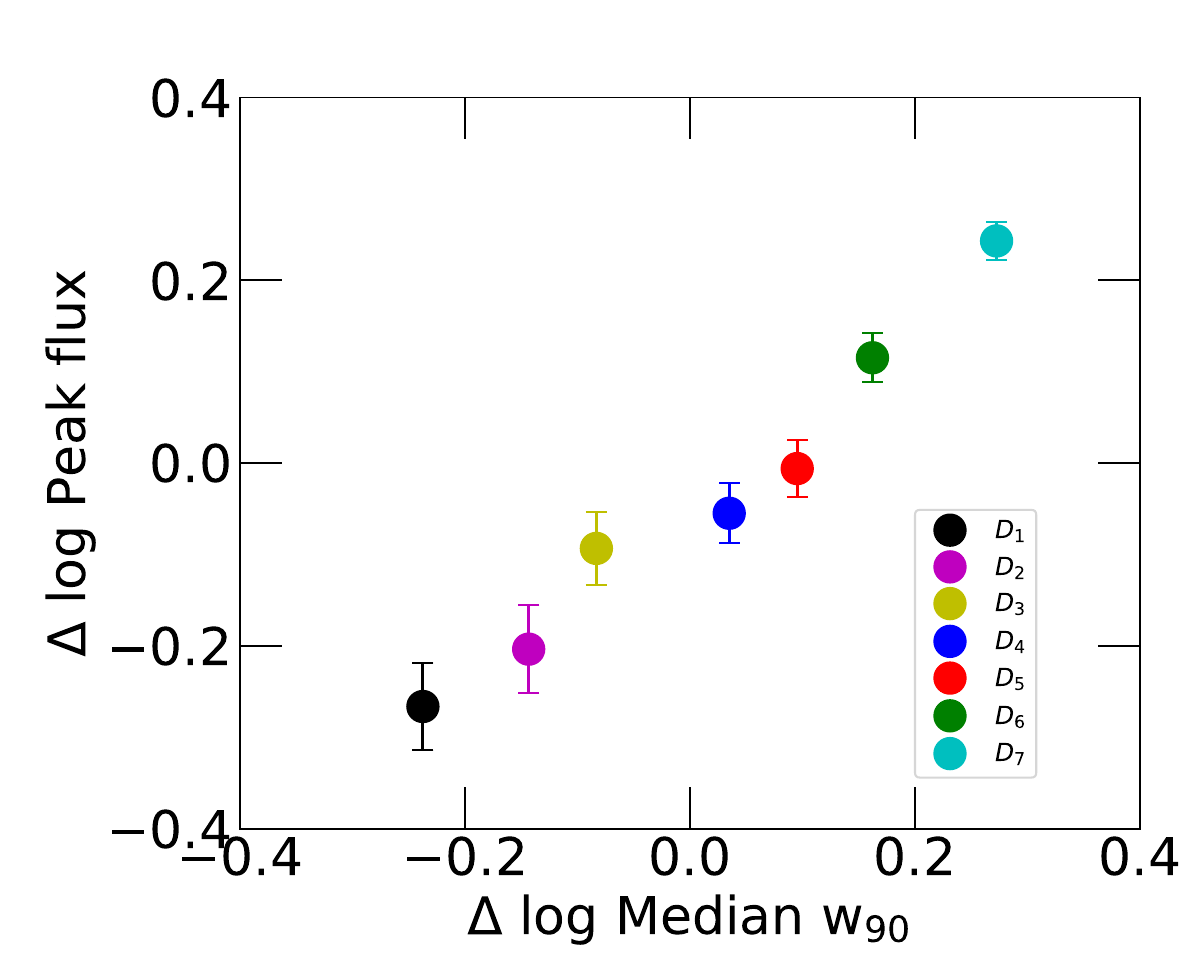}
	\caption{$Left$: The average 3 GHz radio flux (obtained by stacking VLASS images) of SDSS DR16 RQQs versus [O III] $w_{\rm 90}$, for seven bins of [O III] $w_{\rm 90}$. Solid circles: subsamples divided according to [O III] $w_{\rm 90}$. Open circle: the corresponding control sample with matched redshift, black hole mass, luminosity and Eddington ratio. $Right$: The difference plots for the studied bins and their control sample in radio emission and $w_{\rm 90}$.
 Clearly, there exists a strong and intrinsic (i.e., free from possible effects of redshift black hole mass, luminosity and Eddington ratio) correlation between radio emission and [O III] $w_{\rm 90}$ in RQQs.}
\label{fig:radiovsw90}
\end{figure*}

The median-stacked 3 GHz flux densities of each subsample are plotted in the left panel of Fig. \ref{fig:radiovsw90}. It is evident that the stacked radio emission of our subsamples strongly correlates with $w_{\rm 90}$. Meanwhile, their control samples show rather similar median radio fluxes. This indicates that the observed correlation between radio emission and $w_{\rm 90}$ is intrinsic but not due to the effects of fundamental parameters including $z$, $M_{\rm BH}$, $L_{\rm bol}$ and $R_{\rm edd}$ which could affect both radio flux and $w_{\rm 90}$. Furthermore, the radio-loud fraction (based on radio-detected sources) of each subsample and its corresponding control sample show no significant disparity, thus the dependence of median-stacked radio emission on $w_{\rm 90}$ can not be attributed to  the contribution of RLQs either.

\begin{figure*}
	\includegraphics[width=3.4in]{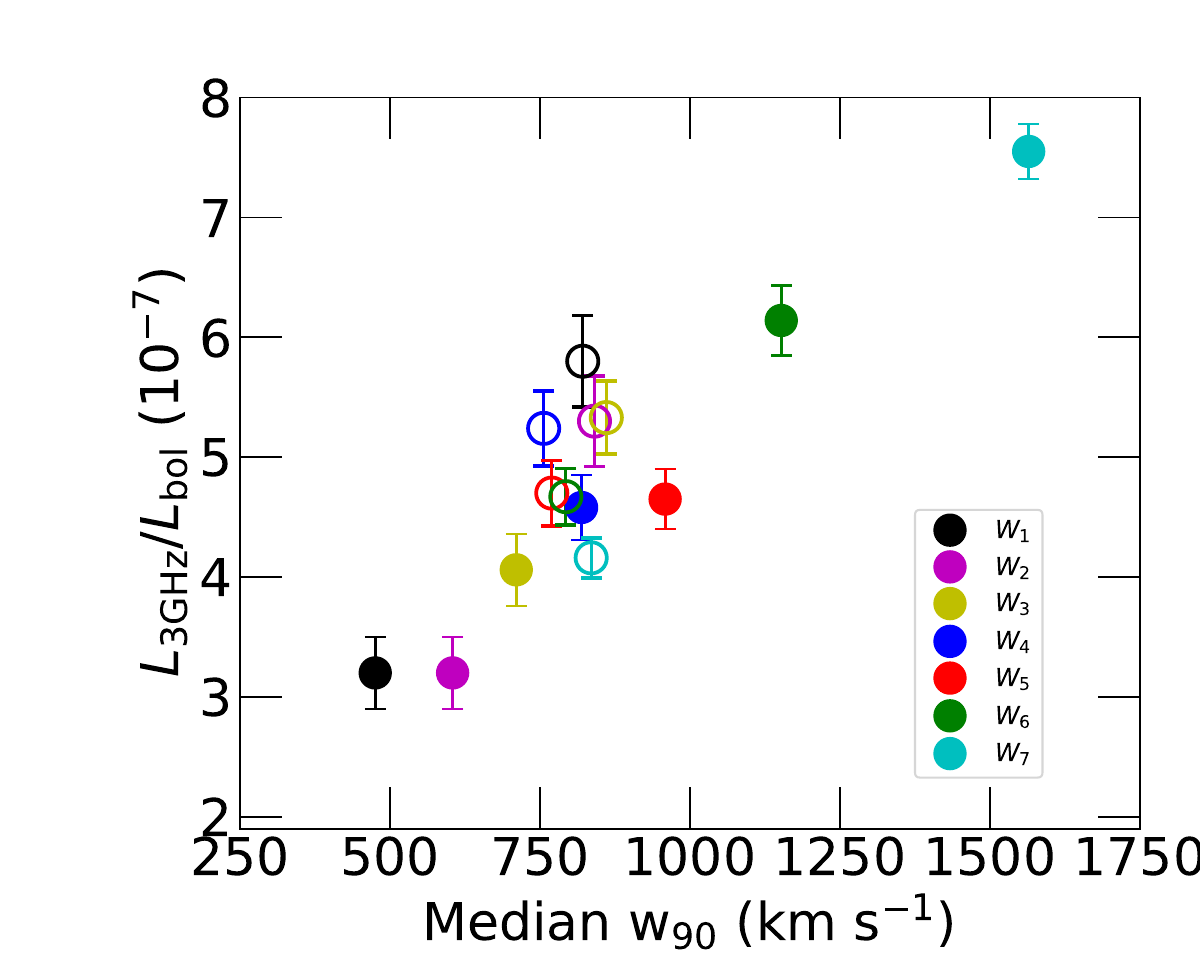}
	\includegraphics[width=3.4in]{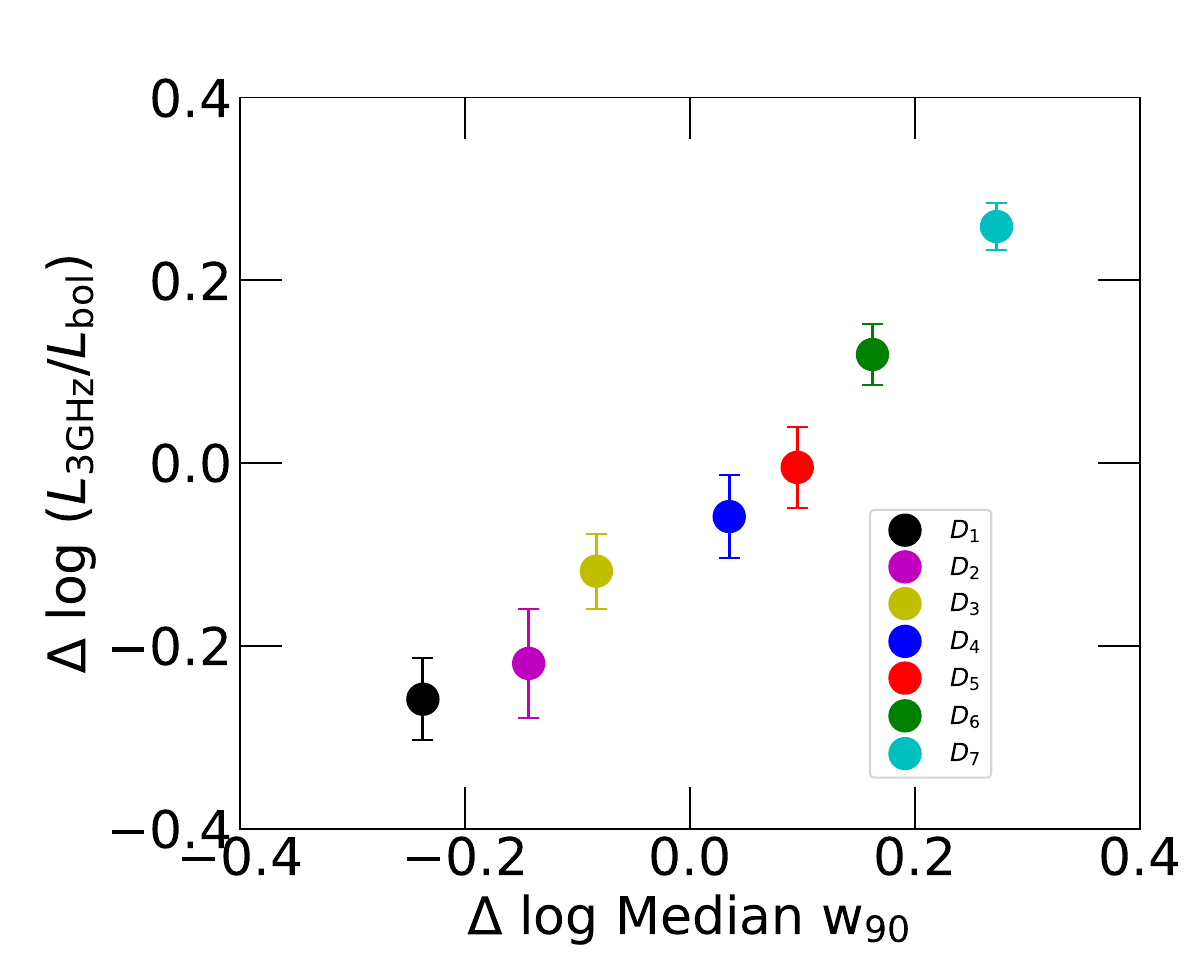}
	\caption{Same as Figure 2 but for $L_{\rm 3GHz}/L_{\rm bol}$ verus [O III] $w_{\rm 90}$.}
\label{fig:Lradiovsw90}
\end{figure*}

To better illustrate the intrinsic link between radio emission and $w_{\rm 90}$, 
we plot the difference of the median radio flux between each subsample and its control sample, versus the difference in the median $w_{\rm 90}$, in the right panel of Fig. \ref{fig:radiovsw90}.
For convenience, we use $D_{1}$ to $D_{7}$ to denote the subsamples in this difference space.
As the observed radio flux of an individual quasar could be highly dependent of its redshift and bolometric luminosity, we also perform analyses through stacking $L_{\rm 3GHz}/L_{\rm bol}$ (instead of directly stacking the radio image), where $L_{\rm 3GHz}$ is calculated by multiplying the radio flux (with k-correction considered assuming a spectral index of 0.5 ($f_{\rm \nu} \propto \nu^{-\alpha}$) ) measured in each pixel in the VLASS cutout image and 4$\pi$$D^{2}$. The corresponding results are shown in Fig. \ref{fig:Lradiovsw90}. A simple linear trend is shown in the right panel of Fig. \ref{fig:radiovsw90} and \ref{fig:Lradiovsw90}, suggesting a single mechanism may possibly dominate the intrinsic positive radio emission -- $w_{\rm 90}$ relation we find.

\section{Discussion}

Our results presented in Fig. \ref{fig:radiovsw90} confirm the positive correlation between radio emission and $w_{\rm 90}$ reported by \citetalias{ZZ14}, but in type 1 RQQs instead of obscured quasars.
The much larger sample of our type 1 quasars enables us to perform radio image stacking to derive median radio fluxes for subsamples of RQQs as a function of $w_{\rm 90}$. More importantly, we find the radio -- $w_{\rm 90}$ relation remains after controlling the effect of $z$, $M_{\rm BH}$, $L_{\rm bol}$, and $R_{\rm edd}$. This indicates the observed radio -- $w_{\rm 90}$ relation is intrinsic, instead of driven by other fundamental physical parameters. 
We note the radio -- $w_{\rm 90}$ correlation slope we observed is steeper (see Fig. \ref{fig:radiovsw90}) than that given by \citetalias{ZZ14}. This is because the radio non-detected individual sources in \citetalias{ZZ14} have preferentially smaller $w_{\rm 90}$, and placing upper limits to their radio emission would weaken and flatten the correlation.
The cons of such process (i.e., using radio flux upper limits for non-detected sources) are avoided in this work through image stacking analyses.



The significant intrinsic correlation between radio emission and [O III] $w_{\rm 90}$ found in our work provides strong evidence that the commonly observed AGN outflows in NLR could indeed make an important contribution to radio emission in general type 1 RQQs, i.e., the synchrotron emission may be related with shock acceleration driven by outflows as proposed in \citetalias{ZZ14}. As seen in Fig. \ref{fig:radiovsw90}, the proximate linear trend in the plane of radio emission and $w_{\rm 90}$ 
suggests that a single mechanism is dominating the correlation.
Contrarily, non-linear correlation between radio power and other properties could suggest multiple origins of radio emission \citep[e.g.,][]{2021AJ....162..270R}.

We note, 
while a number of works \citep[e.g.,][]{2013MNRAS.433..622M,ZZ14,2018MNRAS.477..830H, 2019A&A...631A.132M,2021MNRAS.500.2871S} has reported strong correlations between radio power and [OIII] width/outflow velocity, whether some of the correlations are intrinsic have been questioned \citep{2016ApJ...817..108W,2018ApJ...865....5R,2023ApJ...954...27A}.
As discussed in \cite{2016ApJ...817..108W,2018ApJ...865....5R,2023ApJ...954...27A}, 
both radio emission and [OIII] width depend on the mass of host galaxy that more massive host could produce stronger radio emission and  wider [OIII]. The correlation between radio emission and [OIII] width would become much weaker or disappear, after excluding the effect of the host galaxy gravitational potential on the [OIII] line width (i.e., normalize the [OIII] dispersion $\sigma$ to either the stellar velocity dispersion or the stellar mass) \citep{2016ApJ...817..108W,2018ApJ...865....5R,2023ApJ...954...27A}. 

In this work, we have controlled the effects of SMBH mass, luminosity Eddington ratio and redshift, thus the effect of host gravitational potential could also have been corrected considering the well established M-$\sigma*$ relation. 
Therefore, 
our results indicate the existence of an intrinsic and direct correlation between the radio emission and [OIII] outflow width for RQQs. The discrepancy between this work and \cite{2016ApJ...817..108W,2018ApJ...865....5R} could be due to 1) we focus on quasars while their samples are predominantly less luminous, 2) we adopt $w90$ to quantify the [OIII] outflow width while they adopted [OIII] line dispersion $\sigma$, and 3) we stack radio images to include individually radio non-detected sources and represent the whole RQQ population, while they only utilized a small fraction of their samples with radio detection which may suffer bias due to radio detection incompleteness. 

As the corona contribution to radio emission in RQQs is expected to be weak at 1.4 or 3 GHz (\citealt{2022MNRAS.512..296L}, though may be more significant at high frequencies \citealt{2022MNRAS.510.1043B}), it is neglected hereafter.
Below we discuss possible origins for the observed radio emission --  $w_{\rm 90}$ relation, including  quasar wind, jet, and star formation in host galaxies. 


\subsection{Quasar-winds driven NLR outflow}
It is generally believed that AGN disk-winds driven by magnetic fields and/or thermal expansion and/or radiation pressure \citep{2007ASPC..373..267P}, are prevalent in luminous (evaluated via $L_{\rm bol}$ or $R_{\rm edd}$, i.e., $R_{\rm edd}$ $\sim$ 0.01-1, \citealt{2012ARA&A..50..455F}) AGN and quasars. Such winds interact with the gas surrounding the AGN, including the ISM of the host galaxies, and produce the observed ionized gas outflows including the broad and blueshifted X-ray and UV absorption lines, and optical emission lines (e.g., [O III] in NLR).  
More recently, \cite{2023ApJ...943...98M} find the winds driven by radiation pressure is the dominant driving mechanism behind the NLR [O III] outflows for nearby AGN with $L_{\rm bol}$ > $10^{44}~\rm erg~s^{-1}$, through finding a strong correlation between maximum outflow launch radii from their models, with that derived from the high-resolution optical observations. 
 
Given the high luminosity of our sample (99\% with log $L_{\rm bol}$ > $44~\rm erg~s^{-1}$, 98\% with log $R_{\rm edd}$ > $-$2, and the average log $L_{\rm bol}$ for our seven subsamples ranging from 45.19 -- 45.63$~\rm erg~s^{-1}$), the wide-angle
quasar winds are speculated be strong and efficient in driving [O III] outflows responsible for the shock-driven radio emission in our RQQs. 
Note \citetalias{ZZ14} suggested a threshold of 45.5 ($\pm0.4$ dex) on log $L_{\rm bol}$ for wind driving NLR outflow.
Using the $M_{\rm BH}$ - $\sigma_{*}$ relation \citep{2013IAUS..295..241K}, we can roughly estimate the expected $w_{\rm 90}$ (= 3.29 $\sigma_{*}$) for our quasars if the NLR kinematics 
is dominated by the gravitational potential of host galaxy \citep{2005ApJ...627..721G}.
We find the expected average $w_{\rm 90}$ dominated by host gravity is 565 and 616 $\rm km~s^{-1}$ in $W_{\rm 1}$ and $W_{\rm 2}$ and not exceed 700 $\rm km~s^{-1}$ in other subsamples (converted from the median $M_{\rm BH}$ for each subsample). This indicates that there is likely no outflow signature in $W_{\rm 1}$ and $W_{\rm 2}$ as their observed $w_{\rm 90}$ are either less or consistent with host-dominated, while for $W_{\rm 3}$ to $W_{\rm 7}$, their observed $w_{\rm 90}$ are considerably larger and could well be dominated by the quasar driven outflows\footnote{The median width of $w_{\rm 90}$ for starburst-driven winds is about 600 $\rm km~s^{-1}$ \citep{2014MNRAS.439.2701H}.}. 
Though the observed small $w_{\rm 90}$ in W1 and W2 suggests weak or no outflows in these two subsamples, 
it does not mean that the central engine at their luminosity is not able to drive the winds, as their control samples with matched luminosity and SMBH mass have $w_{\rm 90}$ significantly larger than the expected ones driven by the host (see Fig. \ref{fig:radiovsw90}). This can be explained that the presence of outflows depends on, not only the luminosity/accretion, but also the existence of dense material for the winds to run into \citep{2021AJ....162..270R}. 

We evaluate the conversion efficiency $\eta$ (i.e., $L_{\rm wind}$ = $\eta$ $\times$ $L_{\rm bol}$) required to explain the observed $L_{\rm radio}$ 
($L_{\rm radio}$ = $\eta^{\prime}$ $\times$ $L_{\rm wind}$). Following \citetalias{ZZ14} and \cite{2018MNRAS.477..830H},
we adopt $\eta^{\prime}$ = 3.6 $\times$ $10^{-5}$ as the conversion efficiency of wind kinetic energy into radio luminosity  
The observed $L_{\rm radio}$,
ranging from 2.8 $\times$ 10$^{38}$ to 3.3 $\times$ 10$^{39}$ $~\rm erg~s^{-1}$, are obtained from median stacking of our subsamples. 
We find $\eta$ ranges from 0.5\% to 2.1\% for our subsamples of $W_{3}$ to $W_{7}$ for which the observed $w_{\rm 90}$ should be dominated by [O III] outflows. These values of $\eta$ are well consistent with the ones reported in previous works on quasar powered winds/outflows probed with ionized gas \citep[e.g.,][] {2017ApJ...850...40R} and also the efficiencies assumed in numerical simulations \citep[e.g.,][]{2016MNRAS.458..816H}. Therefore, the observed median-stacked radio emission of our RQQ samples is also quantitatively in accordance with the wind-shock origin.


\subsection{Jet contribution}
Alternatively, low-power jets in RQQs may also, through interacting with the surrounding materials, establish the correlation between [O III] gas kinematics and radio emission. The radio jet co-spatial or aligned with the outflowing ionized gas could provide strong support to this scenario.
In radio-loud AGN, numerous previous studies have shown radio jets could drive the multi-phase (including [OIII]) outflows through interaction with ISM \citep[e.g.,][]{1999MNRAS.307...24V,2006ApJ...650..693N,2008ApJS..175..423P,2008A&A...491..407N,2017A&A...600A.121N,2017A&A...599A.123N,2014A&A...565A..46D,2008MNRAS.387..639H,2013Sci...341.1082M,2013A&A...552L...4M,2019ApJ...879...75H}. The alignments between the kinematics of optical line-emitting gas and the morphology of radio jets are widely found \citep[e.g.,][]{2008ApJS..175..423P}, where the alignments preferentially appear strong in compact radio-loud sources with jet size less than 20 kpc whose compact radio jets could strongly perturb the ISM \citep[e.g.,][]{2021A&ARv..29....3O}.

The correlation between kinematics of optical line-emitting gas and the radio jets' morphology is also found in some RQ AGN and RQQs \citep[e.g.,][]{2019A&A...627A..53H,2019MNRAS.485.2710J,2021MNRAS.503.1780J,2021A&A...648A..17V,2022MNRAS.512.1608G}. For instance, \cite{2019MNRAS.485.2710J} found the highly perturbed [OIII] ionized gas spatially coincides with the radio jets in 10 nearby RQQs with $z <$ 0.2, where the extremely wide [OIII] widths characterized by $w_{\rm 80}$ ($\sim$ 900$-$1200 $~\rm km~s^{-1}$) are detected  in the vicinity of the radio jets.
In addition, if the compact radio emission in RQQs (e.g., $<$ 10 kpc in 86\% RQQs in \citealt{2021MNRAS.503.1780J}, and $<$ 5 kpc for all from \citetalias{ZZ14}) could be dominantly attributed to weak-jets, likely the power scaled-down version of the jets in compact and unresolved radio-loud AGN, one may speculate that these weak-jets in RQQs could similarly correlate with the kinematics of NLR gas. 

However, are the jets are ubiquitous in RQQs? On the one hand, based on high-resolution radio observations (VLA and Merlin), jet-like linear radio structures are seen in about 30-50$\%$ of nearby RQ AGN, suggesting jets are likely common in RQ AGN \citep{2023Galax..11...85S}. Meanwhile at present only a small fraction
RQQs are found to have linear jet-like traits (e.g., 1/7 from \citealt{2010AJ....139.1089L} and 1/3 from \citealt{2021MNRAS.503.1780J}) by the high resolution observations. Particularly, \cite{2021MNRAS.503.1780J} found that, for 2/3 of their [O III] width (400-1600 $\rm km~s^{-1}$) selected RQQ sample, the origin of the radio emission is uncertain and could be quasar-driven winds instead of weak-jets. Note the median radio power of the RQQs in \cite{2021MNRAS.503.1780J}\footnote{We convert the 1.4 GHz emission of \cite{2021MNRAS.503.1780J} to 3GHz (to compare with our results), assuming a radio spectra index of -0.5.} is 6 $\times$ 10$^{39}$ $~\rm erg~s^{-1}$, which is even higher than our subsamples (ranging from 2.8 $\times$ 10$^{38}$ to 3.3 $\times$ 10$^{39}$ $~\rm erg~s^{-1}$). 
This suggests the radio emission is our RQQs is unlikely dominated by jets. 
On the other hand, recently \cite{2019A&A...622A..11G} and \cite{2021MNRAS.506.5888M} found the distribution of radio emission is not bimodal by using the data from Low Frequency Array (LOFAR) Two-Metre Sky Survey \citep[LoTSS,][]{2019A&A...622A...1S} for both detected and non-detected quasars, suggesting jet launching may operate in all the quasars but with different powering efficiency. In this sense, weak radio jets, scaled down from their extended powerful analogues, may exist universally even in radio faintest objects. Even so, as noted by \citetalias{ZZ14}, most of such jets have to be highly compact \citep[$<$ 1 kpc in $\sim$ 50\% RQQs of][]{2021MNRAS.503.1780J}, and consequently we would face the problem that how the highly compact jets work to power energetic galaxy-wide outflows at larger spatial scales \citep[can be out to 10 kpc;][] {2013MNRAS.436.2576L}. Future deep and high-spatial resolution radio observations of a sample of RQQs (with comparable radio power with our sample) could be helpful to further tackle this issue through exploring whether the power of (spatially-resolved) weak jets in RQQs could correlate with [O III] line width (i.e., $w_{90}$), after controlling the effects of fundamental parameters including black mass, luminosity and Eddington ratio.

We conclude that although we can not completely exclude the weak-jets as the radio origin for our [O III] sample, they are more uncertain than the wind-shocked scenario. 
Note the radio emission of our subsample of $W_{1}$ and $W_{2}$ is more likely associated with compact weak-jets or star-formation in the host (instead of wind-driven), as their small [O III] line widths exhibit no evidence of outflow. 
If their radio emission (of $W_{1}$ and $W_{2}$) are actually dominated by weak jets, it may be a hint that while weak-jets are ubiquitous in RQQs, they could not dominantly power the [O III] outflows to produce the linear radio-$w_{90}$ correlation.


\subsection{Star-formation}

\begin{figure*}
	\includegraphics[width=3.3in]{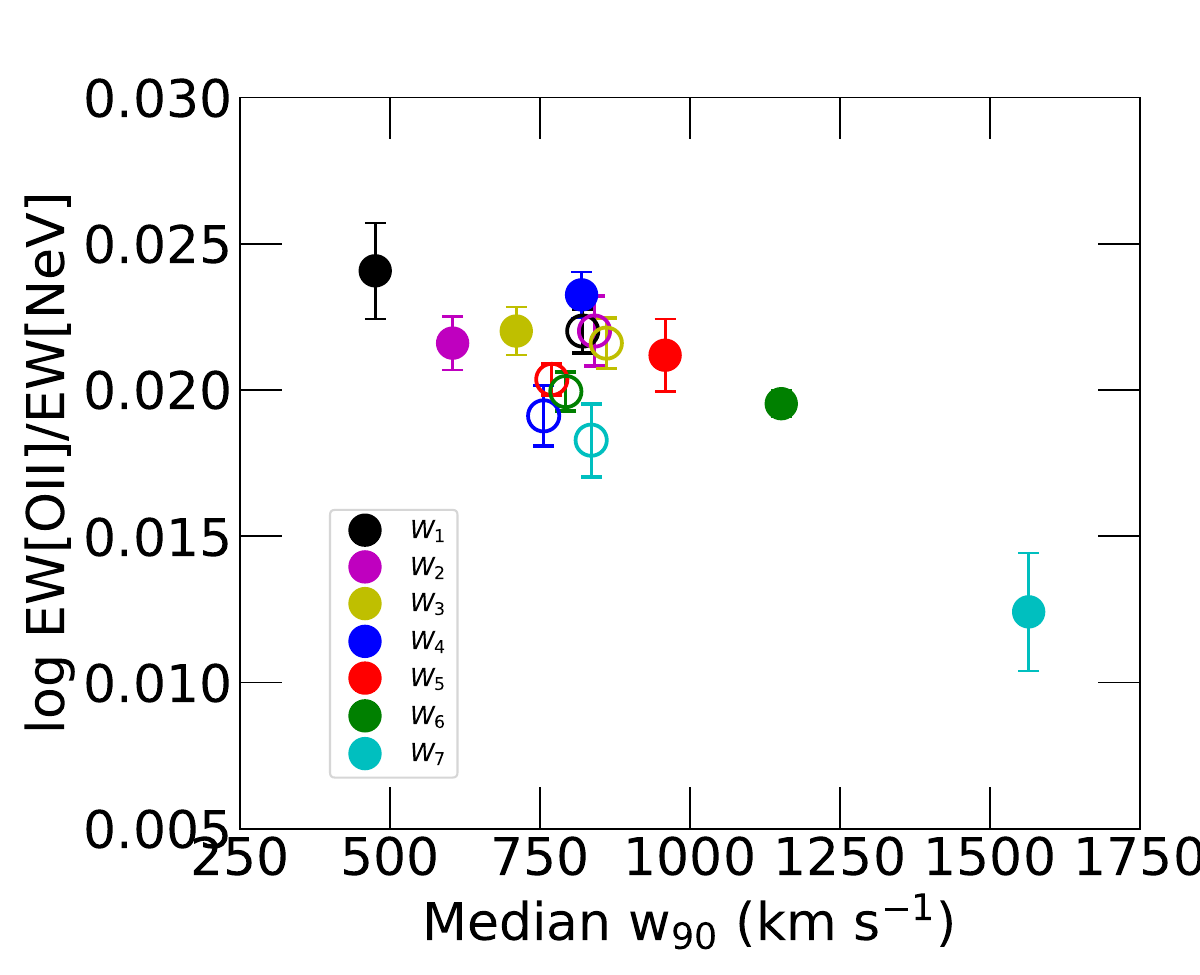}
 	\includegraphics[width=3.3in]{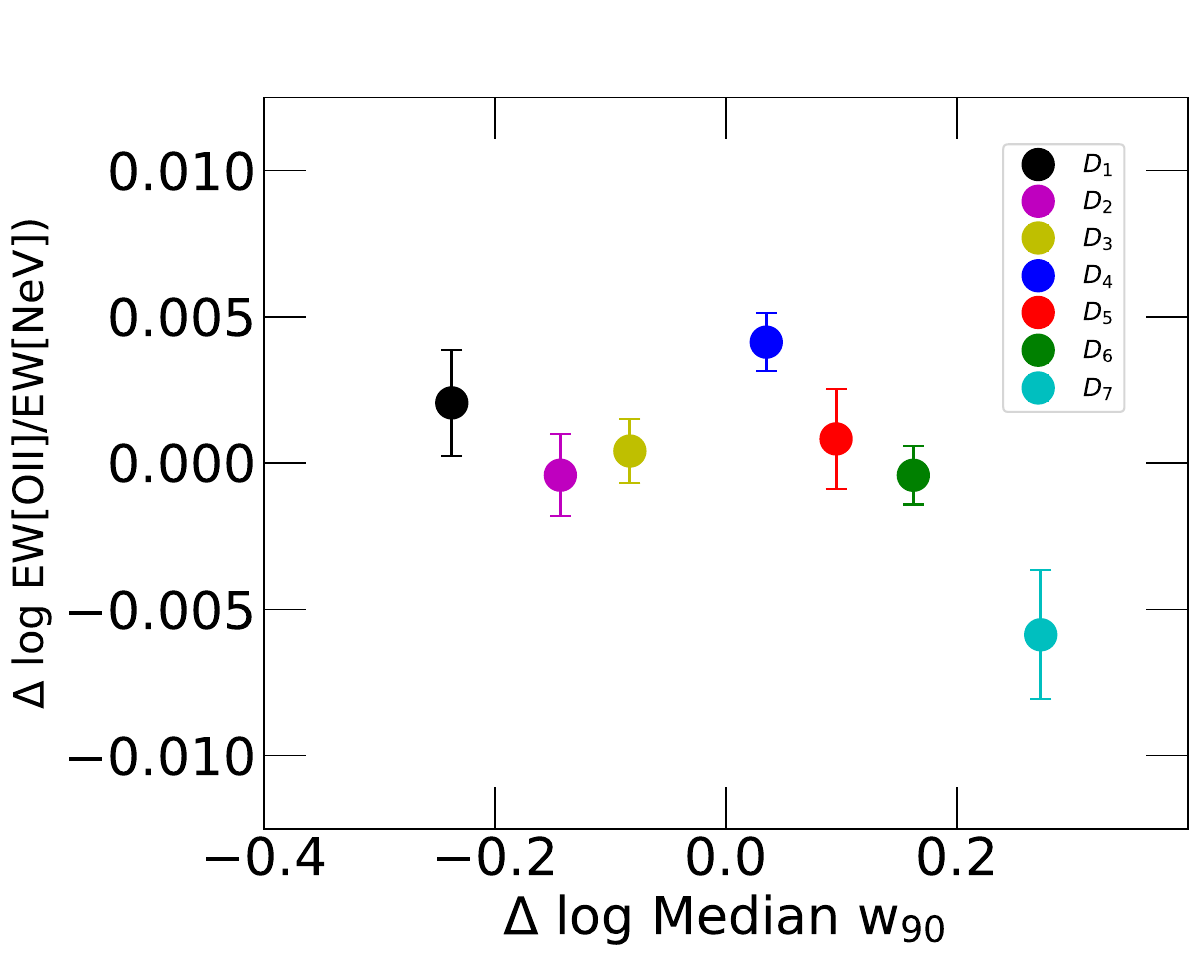}
	\caption{$Left$: The star-formation traced by the flux ratio of [O II]/[Ne V] versus $w_{\rm 90}$ for seven subsamples. Solid circles: subsamples divided according to [O III] $w_{\rm 90}$. Open circle: the corresponding control sample with matched redshift, black hole mass, luminosity and Eddington ratio. $Right$: The difference plots for the studied bins and their control sample in [O II]/[Ne V] and $w_{\rm 90}$.
 \label{star-formation}}
\end{figure*}

The competing contribution to radio emission from the host galaxies needs also to be checked, since the positive feedback \citep[e.g.,][]{2015A&A...582A..63C,2021AJ....162..270R} may enhance the star formation in sources with stronger outflows, thus yield a correlation between $w_{\rm 90}$ and radio emission.
The radio emission from the host can be, either non-thermal synchrotron radiation produced by relativistic electrons accelerated by supernovae \citep[e.g.,][]{2013ApJ...768...37C}, or thermal free-free emission in H II regions ionized by massive stars \citep[e.g.,][]{2019NatAs...3..387P}. These two components are predominant at low and high ($\upsilon >30 \rm GHz$) frequencies, respectively, thus the synchrotron emission is relevant for consideration here. 

Routinely, one may quantify the host contribution to radio emission in AGN from the star formation rate.
However, reliable measurements of SFR are unavailable for most of the individual sources in our large sample.
Instead, we use the line flux ratio of [O II] $\lambda$3729/[Ne V] $\lambda$3426 as a proxy \citep{2022NatAs...6..339C}, to explore the qualitative difference in SFR between various subsamples.
This is because while [Ne V] is dominated by AGN, 
the massive stars in the host galaxies can also excite [O II], thus the relative strength of these two emission lines can represent the star formation activity in quasars.

We median stack the SDSS spectra which covers both [O II] and [Ne V], for our subsamples and their corresponding control sample.
The fluxes of [O II] and [Ne V] are measured through integrating the emission line profiles in the composite spectra normalized by the best-fit continuum model (power-law plus Fe II template). 
The uncertainties associated with the ratio of [O II] to [Ne V] are estimated through bootstrapping the corresponding sample (repeating 100 times).

As shown in the left panel of Fig. \ref{star-formation}, our subsamples exhibit weaker [O II]/[Ne V] at higher $w_{\rm 90}$, suggesting relatively weaker star formation in RQQs with higher $w_{\rm 90}$.
However, after controlling the effect of black hole mass, Eddington ratio and redshift, the dependence of [O II]/[Ne V] on $w_{\rm 90}$ disappears (right panel of Fig. \ref{star-formation}),
and only $W_{\rm 7}$ has tentatively weaker SFR (2.4 $\sigma$) compared with $MW_{\rm 7}$.
This indicates that the higher radio fluxes in subsamples with larger $w_{\rm 90}$ can not be attributed to the hypothetical stronger star formation enhanced by outflows, thus the radio emission in our subsamples is unlikely dominated by the host.
This is in agreement with \cite{2016MNRAS.455.4191Z}, which found in RQQs (Type I and Type II) with $L_{\rm bol}$ $\ge$ 45 $\rm erg~s^{-1}$,
the SF-related radio emission calculated from far-infrared 160 $\rm um$, are insufficient to explain the observed radio emission by about an order of magnitude.


Finally, we stress that the dependence of [O II]/[Ne V] on $w_{\rm 90}$ we see in the left panel of Fig. \ref{star-formation} should not be considered as evidence of negative feedback of [O III] outflows, as the dependence disappears after controlling the effect of SMBH mass, luminosity or Eddington ratio.
Contrarily, we conclude that we see no clear evidence of [O III] outflow feedback, neither positive nor negative, via comparing the [O II]/[Ne V] of our samples with their control samples, while in the subsample of W$_7$ we see tentative negative feedback (weaker SF compare with the control sample).

\section*{Acknowledgements}
We thank the anonymous referee for constructive suggestions which is great helpful in improving the manuscript.
This work was supported by the National Science Foundation of China (Grant Nos: 12033006 and 12192221), and the Cyrus Chun Ying Tang Foundations. ML is supported by the International Partnership Program of Chinese Academy of Sciences, Program No.114A11KYSB20210010. MHZ is supported by the Natural Science Foundation of Jiangxi (No. 20232BAB211024), and the Doctoral Scientific Research Foundation of Shangrao Normal University (Grant No. K6000449)

Funding for the Sloan Digital Sky Survey IV has been provided by the Alfred P. Sloan Foundation, the U.S. Department of Energy Office of Science, and the Participating Institutions. 
SDSS-IV acknowledges support and resources from the Center for High Performance Computing  at the University of Utah. The SDSS website is www.sdss.org. 

The National Radio Astronomy Observatory is a facility of the National Science Foundation operated under cooperative agreement by Associated Universities, Inc. CIRADA is funded by a grant from the Canada Foundation for Innovation 2017 Innovation Fund (Project 35999), as well as by the Provinces of Ontario, British Columbia, Alberta, Manitoba and Quebec. This research has made use of the CIRADA cutout service at URL cutouts.cirada.ca, operated by the Canadian Initiative for Radio Astronomy Data Analysis (CIRADA). CIRADA is funded by a grant from the Canada Foundation for Innovation 2017 Innovation Fund (Project 35999), as well as by the Provinces of Ontario, British Columbia, Alberta, Manitoba and Quebec, in collaboration with the National Research Council of Canada, the US National Radio Astronomy Observatory and Australia’s Commonwealth Scientific and Industrial Research Organisation.

\section*{Data Availability}
This work is based on public images available from the VLASS survey and optical spectra from the SDSS.

\nocite{*}
\bibliographystyle{mnras}
\bibliography{ms} 




\bsp	
\label{lastpage}
\end{document}